# Alternative Anti-Quantum-Channel Disturbance Decoder for Quantum Key Distribution

HUAXING XU,* SHAOHUA WANG, YANG HUANG, YAQI SONG AND CHANGLEI WANG

*China Academy of Electronics and Information Technology, Beijing 100041, China*
*Corresponding author: andy_huaxingxu@126.com*

**In the letter, we propose an anti-quantum-channel disturbance decoder for quantum key distribution (QKD). The decoder is based on a quarter-wave plate reflector-Michelson (Q-M) interferometer, with which the QKD system can be free of polarization disturbance caused by quantum channel and optical devices in the system. The theoretical analysis indicates that the Q-M interferometer is immune to polarization-induced signal fading and its corresponding operator is Pauli Matrix $\sigma_2$, which naturally satisfies the anti-disturbance condition. An Q-M interferometer based time-bin phase encoding QKD setup is demonstrated and the experimental results show that the QKD setup works stably with a low quantum bit error rate and an average safe key rate about 7.34 kbps over 50.4 km standard optical fiber for two hours in the lab.**

Quantum key distribution (QKD) [1] allows two distant participants Alice and Bob to securely share a long random string often called cryptographic keys even in the presence of an eavesdropper Eve. The keys can be used to carry out perfectly secure communication via one-time-pad and perfectly secure authentication via Wigman-Carter authentication scheme [2]. In contrast to conventional cryptography which is mainly based on unproven computational hardness, QKD theoretically offers unconditional security based on fundamental laws of quantum physics, particularly Heisenberg uncertainty principle, to do what mathematics alone cannot do [3]. The first and the best-known QKD protocol is BB84 proposed by C.H. Bennett and G. Brassard in 1984 [4]. Since then, significant progresses have been made in QKD technologies. The unconditional security of QKD has been proved through a series of outstanding works [5–7]. Practical safety of QKD has also been fully studied, such as the decoy state method [8–10] for beating the photon-number-splitting attack which now has been applied in common QKD systems, and the measurement device-independent QKD for removing detector side channel attacks [11]. Up to now, the transmission distance reaches 421 km in optical fiber [12], and 1200 km in free space from the Micius satellite to the Xinglong ground station [13]. Several QKD network testbeds have been built and demonstrated in metro areas [14–17]. Scientists hope to build a global quantum network through quantum satellites connecting terrestrial quantum networks over commercial optical fiber in the future.

To build terrestrial QKD networks over commercial optical fiber, the stability of QKD systems or exactly anti-quantum-channel disturbance is especially crucial and has received extensive attention from both scientific researchers and engineers. Polarization encoding QKD systems rely on complicated feedback compensation because polarization states of photons can be randomly disturbed in optical fiber quantum channel due to environmental vibration and/or temperature variation, and the feedback compensation scheme is not suitable for strong environmental disturbance. Since the phase information encoded in quantum states can be maintained in environmental disturbance, phase encoding or time-bin phase encoding QKD systems are more competitive over overhead and tube optical cable along roads or bridges. However, decoding the phase information with unbalanced-arm Mach-Zehnder or Michelson interferometers will also suffer from polarization disturbance in transmission fiber and result in the fringe visibility of the interferometers varying fast [18]. To solve the problem, A. Muller et al. proposed the "plug and play" bidirectional QKD system which can automatically compensate the polarization disturbance besides the phase drifting in the channel [19], and has been applied in QKD products by ID Quantique, Inc., a well-known Swiss quantum company. Moreover, X. F. Mo et al. proposed the Faraday-Michelson (F-M) unidirectional QKD system [20], which has been applied in the phase encoding QKD products by Anhui Asky Quantum Technology Co., Ltd., and in the latest timing-bin phase QKD products by QuantumCTek Co., Ltd, in China. According to Ref [18], for phase encoding QKD systems the disturbances of quantum channel will be collected in the system if there is polarization-induced fading at the receiver's interferometer. So it is crucial to construct an unbalanced-arm interferometer which can be self-compensating quantum channel disturbance. Up to now, the effective and widely used solutions are mainly based on the two previous schemes or their variants. Both the two solutions are based on the same principle in essence because they both take advantage of the Faraday magneto-optic effect.

In the letter, we propose a completely new scheme as far as we know, which is free of polarization disturbance caused by quantum channel and optical devices in QKD systems by using quarter-wave plate reflector-Michelson (Q-M) interferometers. The physical and theoretical analysis is given and the conclusions reveal that our scheme can automatically eliminate the polarization-induced signal fading. We set up a Q-M interferometer based time-bin phase encoding QKD system, and the system exhibits high degree of stability and low quantum bit error rate (QBER) with the average safe key rate around 7.34 kbps over 50.4 km optical fiber in two hours. Moreover, our scheme can be expected to realize optical integration as shown in our subsequent work [21] and can be applied in magnetic environment because it is without Faraday components.

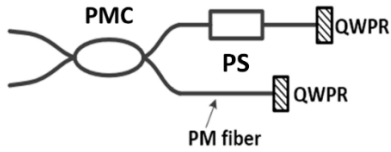

Fig. 1. Unbalanced-arm Michelson interferometer with two quarter-wave plate reflectors (QWPRs) as mirrors, a polarization maintaining coupler (PMC), a phase shifter (PS) and PM optical fibers.

According to Ref. [18], the anti-disturbance condition in an unidirectional phase encoding or time-bin phase encoding QKD system is $L \cdot S = I$ or $L = S$, where $L$ and $S$ represent the operators of the whole long and short arms of the unbalanced-arm interferometer, respectively, and they are unitary. In our scheme we propose the Q-M interferometer, which is composed of a polarization maintaining coupler (PMC) and two unbalanced arms (the upper and lower arms) as shown in Fig.1. Both the upper and lower arms are comprised of polarization maintaining (PM) optical fiber, a quarter-wave plate (QWP) and a reflector. The QWP and reflector can be fabricated into an integral optical component, i.e. a quarter-wave plate reflector (QWPR). The slow and fast axes of the QWP are along $x$ and $y$ direction and the slow and fast axes of the PM optical fiber are along $X$ and $Y$ direction, respectively. The angle between the slow axes of PM optical fiber and QWP is 45 degrees. Besides, there is a phase shifter (PS, as shown in Fig.1) and/or phase modulator (not shown in Fig.1) in one arm, for example in the upper arm, for compensating phase drifting and/or implementing phase encoding. Through the analysis below, we find that either arm of the Q-M interferometer corresponds to Pauli Matrix $\sigma_2$ while that of the Faraday Michelson interferometer corresponds to Pauli Matrix $\sigma_3$. It is easy to confirm that the three Pauli Matrices including $\sigma_2$ and $\sigma_3$ are naturally satisfied the anti-disturbance condition and make QKD systems immune to polarization-induced signal fading. In the letter the mathematical notation is the same as Ref. [22].

Physically, a QWPR can turn $X$-direction linear polarization light to $Y$-direction, and vice versa, when the angle between the polarization direction of the linear light and the slow axis of the QWP equals 45 degrees. As shown in Fig.2 (a), a forward $X$ ($Y$) polarization incident light along the slow (fast) axis of PM optical fiber can be transformed into a backward $Y$ ($X$) polarization output light along the fast (slow) axis of PM optical fiber after the reflection by the QWPR. Due to the same phase accumulation during the round-trip transmission, only the exchange of the X- and Y-polarization states happens between the input and output light, namely, the operator of the long and short arms is the matrix $\begin{bmatrix} 0 & 1 \\ 1 & 0 \end{bmatrix}$. Therefore the output polarization state is independent of PM optical fiber in the interferometer arms and can be expressed as the product of the incident polarization state and Pauli Matrix $\sigma_2$ (ignoring the phase factor $i$).

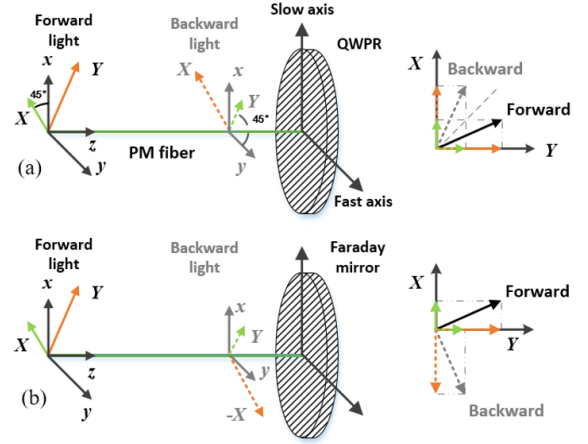

Fig. 2. (a) The forward light (or incident light) and backward light (or output light) after the reflection by a QWPR, where the angle between the direction of $X$ polarization state and the slow axis of the QWP ($x$-direction) is 45 degrees. (b) The forward light (or incident light) and backward light (or output light) after the reflection by a Faraday mirror. The solid and dashed lines respect forward and backward light after the reflection, respectively.

However, for a Faraday mirror, there will be an additional phase $\pi$ besides the polarization state exchange. As shown in Fig.2 (b), the $X$- and $Y$-direction components of forward light (or incident light) correspond to $Y$-direction and $(-X)$-direction components of backward light (or output light) reflected by the Faraday mirror. So the polarization direction of the incident light and the output light after the reflection by the Faraday mirror are always orthogonal with each other and the F-M based QKD scheme will be free of polarization disturbances in the system, as seen in Fig.2 (b). Unlike the Q-M interferometer, the operator of the long and short arms of F-M interferometer is the matrix $\begin{bmatrix} 0 & 1 \\ -1 & 0 \end{bmatrix}$, which correspond to Pauli Matrix $\sigma_3$.

To describe the change of polarization more intuitively, we take the coordinate system as right handed and the light propagation direction as $+z$ direction when the light propagates to the QWPR along PM optical fiber, and the coordinate system is left handed after the reflection by the reflector. For the convenience of theoretical analysis, we use the same notations with those in Ref. [22]. For the Q-M interferometer, the operator of PM optical fiber in the long arm is $l = U(\delta/2, 1, 0, 0)$, where $\delta$ is the birefringence strength of the PM fiber, and the operator of the QWPR is $QR_{\lambda/4} = U(\pi/4, 0, 1, 0)^t \cdot U(\pi/4, 0, 1, 0) = U(\pi/2, 0, 1, 0) = i\sigma_2$, where:

$$U(\gamma, s_1, s_2, s_3) = \sigma_0 \cos\gamma + i(s_1\sigma_1 + s_2\sigma_2 + s_3\sigma_3)\sin\gamma,$$
$$s_1^2 + s_2^2 + s_3^2 = 1$$
$$\sigma_0 = \begin{bmatrix} 1 & 0 \\ 0 & 1 \end{bmatrix}, \quad \sigma_1 = \begin{bmatrix} 1 & 0 \\ 0 & -1 \end{bmatrix},$$
$$\sigma_2 = \begin{bmatrix} 0 & 1 \\ 1 & 0 \end{bmatrix}, \quad \sigma_3 = \begin{bmatrix} 0 & i \\ -i & 0 \end{bmatrix}. \quad (1)$$

The angle $\gamma/2$ corresponds to the birefringence strength of the transmission medium, $\sigma_0$ is 2-D unit matrix, $\sigma_1$, $\sigma_2$, $\sigma_3$ are Pauli matrices, and $s_1$, $s_2$, $s_3$ are Stokes vectors. These Stokes parameters, $s_1$, $s_2$, $s_3$, originate from the X-Y and +45° and -45° components of rectangular birefringence, and from the circular birefringence, respectively. Then the long arm operator $L$ will be:

$$L = \overleftarrow{l} \cdot QR_{\lambda/4} \cdot \overrightarrow{l}$$
$$= U(\frac{\delta}{2},1,0,0)^t \cdot U(\frac{\pi}{2},0,1,0) \cdot U(\frac{\delta}{2},1,0,0)$$
$$= U(\frac{\delta}{2},1,0,0) \cdot U(\frac{\pi}{2},0,1,0) \cdot U(\frac{\delta}{2},1,0,0) \quad (2)$$
$$= U(\frac{\pi}{2},0,1,0) = QR_{\lambda/4} = i\sigma_2.$$

Where the arrows ← and → indicate the backward and forward propagation and the superscript $t$ designates the transposed matrix. For reciprocal optical element, the backward propagation notation equals the transposed matrix of the notation in forward coordinate system. The conclusion also applies to the short arm operator $S$.

Now considering double Q-M interferometers in two distant participants Alice and Bob, respectively, connected with each other by commercial optical fiber quantum channel, there will be two paths for the pulse to interfere [18]:

Path 1: $L_a \to$ channel $\to S_b$;
Path 2: $S_a \to$ channel $\to L_b$;

Where the subscripts $a$ and $b$ represents the operators of the components at the Alice's and Bob's side, respectively. The transformation matrices of the two paths can be described respectively by:

$$P_1: (\overleftarrow{s_b} \cdot QR_{\lambda/4} \cdot \overrightarrow{s_b}) \cdot C \cdot (\overleftarrow{l_a} \cdot QR_{\lambda/4} \cdot \overrightarrow{l_a}) e^{i\varphi_a}$$
$$= e^{i\beta_b} \cdot QR_{\lambda/4} \cdot e^{i\phi} \cdot C \cdot e^{i\alpha_a} \cdot QR_{\lambda/4} \cdot e^{i\varphi_a},$$
$$P_2: (\overleftarrow{l_b} \cdot QR_{\lambda/4} \cdot \overrightarrow{l_b}) e^{i\varphi_b} \cdot C \cdot (\overleftarrow{s_a} \cdot QR_{\lambda/4} \cdot \overrightarrow{s_a}) \quad (3)$$
$$= e^{i\alpha_b} \cdot QR_{\lambda/4} \cdot e^{i\phi} \cdot C \cdot e^{i\beta_a} \cdot QR_{\lambda/4} \cdot e^{i\varphi_b},$$

Where $l_i$ and $s_i$ ($i=a,b$) represent the operators of the long and short PM optical fiber in the Q-M interferometer's arms, $\alpha_i$ and $\beta_i$ are the phase caused by the long and short arm of interferometers, respectively, $\phi$ is the phase of transmission fiber, and $\varphi$ is the phase shift from the phase shifter or phase modulator in the interferometer. Supposing the input Jones vector is $E_{in}$ at Alice's side, the output of Bob's interferometer can be written as:

$$E_{out} = \frac{1}{4}(e^{i(\alpha_a+\beta_b+\varphi_a+\phi)} + e^{i(\alpha_b+\beta_a+\varphi_b+\phi)})QR_{\lambda/4} \cdot C \cdot QR_{\lambda/4} \cdot E_{in}. \quad (4)$$

Where the factor 1/4 originates from the PMCs of Alice's and Bob's interferometers. Considering that $C$ is unitary, the interference output power can be expressed as:

$$P_{out} = E_{out}^+ \cdot E_{out}$$
$$= E_{in}^+ [\frac{1}{4}(e^{i(\alpha_a+\beta_b+\varphi_a+\phi)} + e^{i(\alpha_b+\beta_a+\varphi_b+\phi)})QR_{\lambda/4} \cdot C \cdot QR_{\lambda/4}]^+$$
$$\cdot [\frac{1}{4}(e^{i(\alpha_a+\beta_b+\varphi_a+\phi)} + e^{i(\alpha_b+\beta_a+\varphi_b+\phi)})QR_{\lambda/4} \cdot C \cdot QR_{\lambda/4}] E_{in} \quad (5)$$
$$= \frac{E_{in}^+ \cdot E_{in}}{8}[1+\cos(\Delta\alpha+\Delta\beta+\Delta\varphi)]$$
$$= \frac{P_{in}}{8}[1+\cos(\Delta\alpha+\Delta\beta+\Delta\varphi)].$$

Where $\Delta\alpha=\alpha_a-\alpha_b$, $\Delta\beta=\beta_a-\beta_b$, $\Delta\varphi=\varphi_a-\varphi_b$. This means that the interference output $P_{out}$ is independent of any polarization perturbation in the whole QKD system, especially that caused by the quantum channel. In an ideal case, $\Delta\alpha$ and $\Delta\beta$ are invariable, hence interference fringe is only modulated by the phase shifter or phase modulator in the interferometer. In the real case, the phase drifting of $\Delta\alpha$ and $\Delta\beta$ caused by the fluctuation of temperature or environmental vibration can be solved by active compensation.

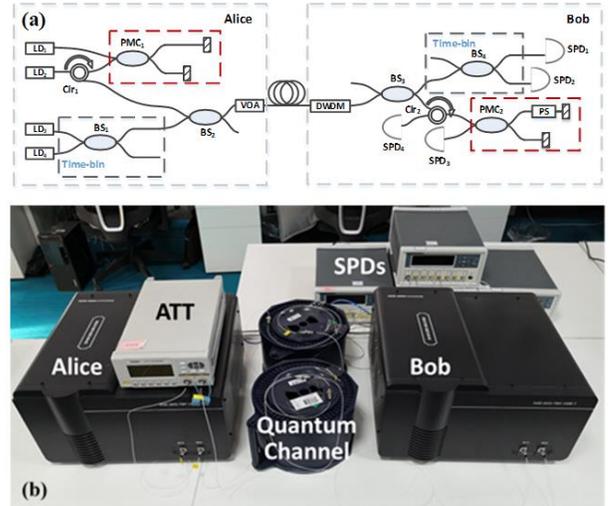

Fig.3 (a) Schematic diagram of Q-M based time-bin phase encoding intrinsic-stabilization QKD system. Q-Ms are shown in the red dashed boxes. $LD_1$-$LD_4$ are lasers, $Cir_1$ and $Cir_2$ are circulators, $PMC_1$ and $PMC_2$ are polarization maintaining couplers, $BS_1$-$BS_4$ are beam splitters, PS is a phase shifter, DWDM is a dense wavelength division multiplexer, VOA is a variable optical attenuator, and $SPD_1$-$SPD_4$ are four avalanche diode single photon detectors. (b) The real experimental setup in the lab, including the quantum channel of 50.4 km optical fiber.

To prove the theory above, a time-bin phase encoding intrinsic-stabilization QKD experimental system is built. The schematic and the real experimental setup are shown in Fig.3. In the setup both the transmitter Alice and receiver Bob contain a Q-M interferometer as the phase encoder and decoder for phases {0, π} encoding and decoding, respectively, as seen in the red dash boxes in Fig.3 (a). The $BS_1$ and $BS_4$ are used for time-bin encoding and decoding as seen in the black dash boxes in Fig.3 (a). A variable optical attenuator (ATT) at Alice's side is used to attenuate the transmitted light to single photon lever, the $BS_3$ at Bob's side is used for passive basis selection, and a dense wavelength division multiplexer (DWDM) used at Bob's side is for spectral filtering to reduce the scattered and background noise. The system works in a way of decoy-state BB84 protocol including vacuum and weak decoy states [23].

In the experiment, the photons are generated by four strongly attenuated 1549.32 nm distributed-feedback pulsed laser diodes with a pulse width of 500 ps and 100 MHz repetition rate. The average photon number is 0.6 per pulse including transmitted signal, decoy state and vacuum state, and the vacuum state is generated by not triggering the lasers. The ratio of signal, decoy and vacuum state numbers is 6:1:1. The quantum fiber channel from Alice to Bob is 50.4 km (about 9.5 dB loss) in the lab, and four avalanche diode single photon detectors are used at Bob's side with a gate width of 1ns.

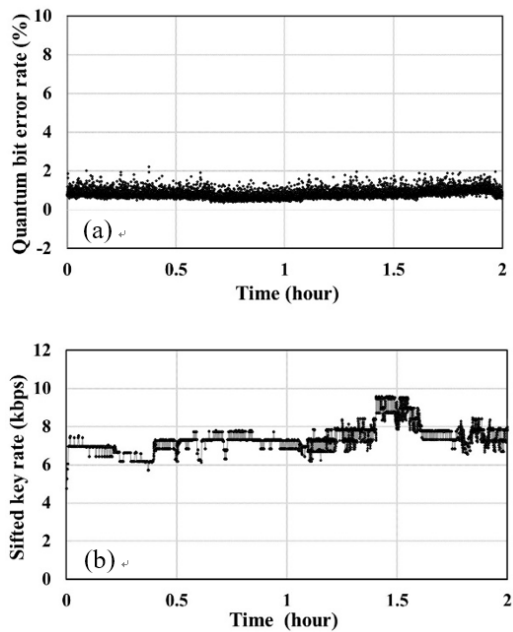

Fig. 4. Temporal fluctuation of the key generation performance. (a) The quantum bit error rate as a function of time. (b) The safe key rate as a function of time, where every point is the average safe key rate in one minute.

With the setup described above, we measured the QBER and safe key rate to examine the performance of the system. As shown in Fig.4, the average QBER and safe key rate are (a) 0.83%±0.23% and (b) 7.34±0.72 kbps during two hours over 50.4 km optical fiber, respectively. Here we deal with the data collected from detectors every second to return a value as a QBER point in Fig. 4 (a). In Fig. 4 (b), the squares represent the average safe keys per second, which is an average of the collected safe keys in one minute. Another external optical attenuator is used to simulate the loss of the optical fiber quantum channel. Assuming the optical fiber loss is 0.2 dB/km, the safe key rate is about 0.25 kbps over 100 km optical fiber (about 19.5 dB total channel loss) under lab condition. The experimental results indicate that the Q-M scheme can keep QKD system working stably.

In summary, we propose an unbalanced-arm Q-M interferometer based QKD scheme, which is free of polarization disturbance caused by optical fiber quantum channel and optical devices in the system. Physical and theoretical analysis has been presented. An experimental verification is also implemented by building a Q-M interferometer based time-bin phase encoding QKD system. The experimental results reveal a long-term low QBER with about 0.83%, and a stable safe key rate with about 7.34 kbps over 50.4 km and 0.25 kbps over equivalent 100 km optical fiber under lab condition. All the components in the Q-M interferometer are conventional commercial passive optical components, and can be easily fabricated. As the Q-M interferometer is without Faraday components, our scheme can be expected to realize optical integration that will be shown in our subsequent work and can be applied in magnetic environment. The theoretical analysis and the experimental results indicate that the Q-M based QKD system can work stably and will be a competitive scheme in practical QKD application.

**Funding.** National Natural Science Foundation of China (No. 61705202); Innovation Funds of China Academy of Electronics and Information Technology.